\documentclass[twocolumn,prb,a4paper]{revtex4}
\usepackage[T1]{fontenc} 
\usepackage[applemac]{inputenc}
\usepackage{graphicx}
\usepackage{dcolumn}
\usepackage{bm}
\usepackage{amsmath,amssymb}
\usepackage{pifont}
\usepackage{graphicx}

\usepackage[usenames,dvipsnames]{color}

\def\PP#1{{#1}}
\def\Ph#1{{#1}}

\def\Os#1{{#1}}
\def\OV#1{{#1}}

\def\rayer#1{\sout{}}
\begin{document}

\graphicspath{{Figures/}}

\title{Birth and growth of cavitation bubbles within \\ water under tension \OV{confined in} a simple synthetic tree}

\author{{Olivier Vincent}}
\author{{Philippe Marmottant}}
\email{philippe.marmottant@ujf-grenoble.fr}
\affiliation{CNRS \& Universit\'e de Grenoble, UMR 5588,  Laboratoire Interdisciplinaire de Physique, Grenoble, F-38041, France}
\author{{Pedro A. Quinto-Su}}
\altaffiliation{Current address: ICN-UNAM, C.P. 04510, M\'exico, D.F.,
M\'exico.}
\author{{Claus-Dieter Ohl}}
\affiliation{Division of Physics and Applied Physics, School of Physical and Mathematical Sciences, Nanyang Technological University, Singapore 637371, Singapore}

\date{\today}

\begin{abstract}
Water under tension, as can be found in several systems including tree vessels, is metastable. Cavitation can spontaneously occur, nucleating a bubble. We investigate the dynamics of spontaneous or triggered cavitation  inside water filled microcavities  of a hydrogel. Results show that a stable bubble is created in only a microsecond timescale, \Os{after transient oscillations}. Then, a diffusion driven expansion leads to filling of the cavity. Analysis reveals that the nucleation of a bubble releases a tension of several tens of MPa, and a simple model captures the different time scales of the expansion process.

\end{abstract}

\maketitle

The dynamics of cavitation is a well studied process in free liquids (cavitation in the wake of propellers for instance),  while cavitation in a microscopic confinement only recently received more attention. One important area of research is the transport of sap in trees and the potential blockage of this transport due to cavitation:  these events can stop circulation of water  in the vessels of real trees\cite{Tyree2003, Cochard2006}, or in synthetic ones that are used for microfluidic flow transport driven by evaporation \cite{Wheeler2008,Noblin2008,Moreau2009}. 

The stability of water at negative pressure, i.e. under tension (see\cite{Herbert2006} for a review), can be tested with hydrogels where the tension is fixed through the relative humidity of the surrounding medium \cite{Wheeler2008,Wheeler2009}. 
However, such a method features only the final equilibrium state, and does not reveal about the cavitation process itself.  The dynamics of cavitation involves fast time scales that are not known, apart from the fact that they are suspected to generate vibrations (noises in trees \cite{Tyree1983}).

In this study we \Os{resolve} the dynamics of  cavitation of water inside hydrogel cavities. Cavitation can occur spontaneously or be triggered by a disturbance created by a laser pulse.  The dynamics of the nucleated bubble and hydrogel cavity are recorded with a high-speed camera or a photodiode. 

\begin{figure}

	\begin{center}
	\includegraphics[scale=1]{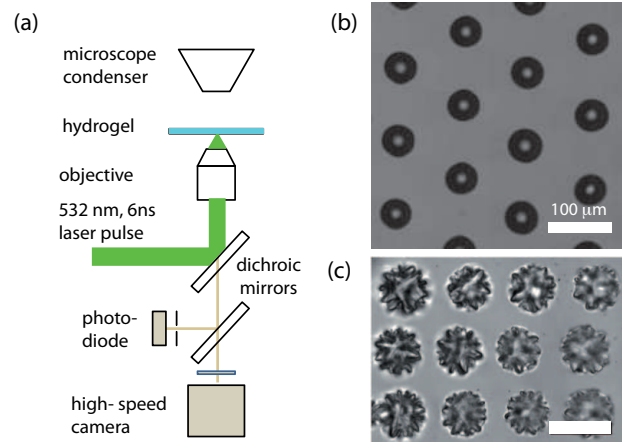}
			\caption{\small (a) Experimental set-up. (b) Microcavities in the hydrogel created by
laser pulses with low UV pre-curing, (c) Microcavities when the hydrogel
is pre-cured for a longer time. }
		\label{fig:Fig1}
	\end{center}
\end{figure}

We use pHEMA hydrogels of the same chemical composition as in\cite{Wheeler2008,Wheeler2009}, but with  a periodic array of holes  or cavities, created using the novel experimental setup shown in Fig. 1a. An Nd:YAG laser ($\lambda =532$ nm, $t_p=6$ ns) is coupled to an inverted microscope. The laser pulses are focused by a 40x water immersion microscope objective inside a thin layer of partially polymerized hydrogel that has been precured with UV light (dimensions $\sim$ 1.5 $\times$1.5 cm$\times$1 mm). The hydrogel sample is bounded by two microscope coverslides separated by PDMS spacers.
A single cavity is created after a laser-induced optical breakdown inside the hydrogel sample.
The system is coupled to an automated translation stage that allows the design of a regular array of holes. 
The shape of the cavities strongly depends on the degree of polymerisation of the hydrogel: while they are smooth and spherical for low pre-polymerisation, they tend to form "cauliflower" shapes when more cured (Fig. \ref{fig:Fig1}b,c), in a manner very similar to what is observed when forcing air in a polyacrylamide gel\cite{Kundu2009}.
Finally, the hydrogel is fully polymerized (Fig. 1b). To fill the the cavities with water, the hydrogel is boiled with water for a few days, \OV{and soaked in degazed DI water}.

\begin{figure}
	\begin{center}
	\includegraphics[scale=1]{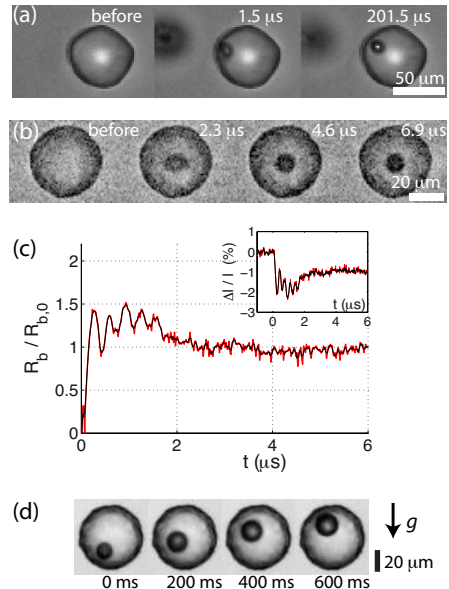}
			\caption{\small Image sequences showing the birth of a cavitation bubble (a) triggered by a laser pulse aimed away from microcavity, here on the left (b) spontaneous without triggering,  here under 40x objective. (c) Photo-diode recording \Os{on a cavity with a radius of 40 $\mu$m:} Estimated \PP{apparent} bubble radius as a function of time, normalized by the bubble \Os{radius} just after nucleation, for a single experiment. \Os{Red : experimental data ; black: slightly filtered signal. }Insert: recorded relative variations of the photodiode signal \OV{from which the radius is estimated}. (d) Long term rise of a bubble in a vertical sample. }
		\label{fig:Fig2}
	\end{center}
\end{figure}

As the hydrogel samples are dried, water is set under tension because of evaporation. 
Cavitation  spontaneously occurs when tension is strong enough, 
or can be directly triggered with laser pulses, focused in the hydrogel, in vicinity of water filled voids.
The gel  is sometimes slightly altered (apparition of grey spot, Fig. \ref{fig:Fig2}a). We speculate that a mild pressure wave is created, travels over the microcavity and thereby increases the liquid tension for a brief moment. 
This mechanical disturbance in the hydrogel is able to trigger cavitation in neighboring microcavities, if  under tension.
The dynamics of cavitation is recorded with a high-speed camera at frame rates of up to 432,000 frames per second, with an exposure time of 1 $\mu$s. In addition,  the central part of the image can be projected simultaneously onto a  photodiode \Os{with a rise time of 0.03 $\mu$s  and bandwidth of 12 MHz, providing a measurement of the mean intensity in order to access faster dynamics.}

\textit{Fast time scales} Our recordings reveal that a single bubble is nucleated, and that its  volume quickly reaches a finite value, \OV{within 1 or 2 microseconds} (Fig. \ref{fig:Fig2}a-c).  
We captured the first image of the nucleated bubble \Ph{after} a delay of 1 $\mu$s after arrival of the laser pulse. If these images are motion blurred due to the exposure time of 1$\mu$s, they show   a transient non-spherical shape, see second image of Fig 2b.
\Os{On the photo-diode signal we record transient high-frequency oscillations at typically several MHz, that we interpret as oscillations of the radius in Fig 2c, with an initial velocity as fast as 30 m/s. We assumed the variation in light intensity to be  proportional to the apparent bubble area on images.} 

The following images of \OV{the high-speed} recording show that the bubble after nucleation reaches a temporary equilibrium radius and then slowly grows. This is in contrast to cavitation nucleation in a free liquid, where the bubble expands to a maximum size and then rapidly collapses due to the pressure of the surrounding liquid. 
Experiments with an observation from the side also evidence that bubbles rise due to their buoyancy (Fig. \ref{fig:Fig2}d), meaning they are not attached by a contact line and are quickly spherical.

\begin{figure}
	\begin{center}
	\includegraphics[scale=1]{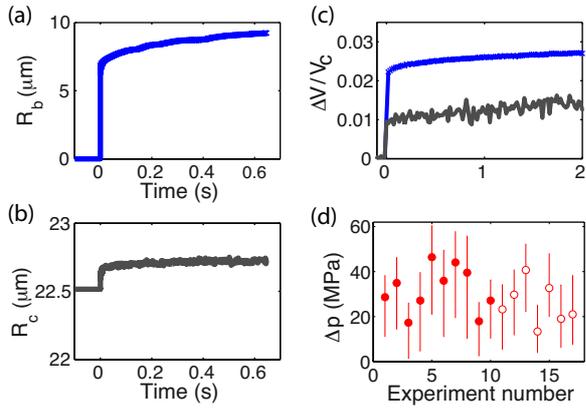}
			\caption{\small Evolution of (a) the bubble radius $R_b$  and (b) the cavity radius $R_c$. (c) We infer that the bubble volume \Ph{increase} is larger than cavity volume increase, meaning that the liquid volume decreases. \PP{(d)} Estimation of the tension that is relaxed, for triggered (full circles) and spontaneous bubbles (open circles). }
		\label{fig:Fig3}
	\end{center}
\end{figure}

The instantaneous volume of the bubble ($V_b$) and of the cavity ($V_c$) are obtained by extracting an effective radius from image analysis (Fig. \ref{fig:Fig3}), consisting in tracing iso-values lines on grey levels bitmaps. 
Interestingly, the volume of liquid \Ph{$V_l=V_c-V_{b,0}$} just after nucleation decreases. This is due to the  compressibility of water, which has an adiabatic bulk modulus
\footnote{At  $20^\circ$C, The adiabatic and isothermal compressibilities of water are very close and differ less than 1 percent. Their variation with pressure is also small: about 2.5 percents each 10 MPa.}
$K_l=2.2$ GPa at $20^\circ$C. Therefore, the variation of liquid volume \Ph{$\Delta V_l$ $(=\Delta V_c-V_{b,0})$} gives information on the pressure change $\Delta p$ in the liquid: 

\begin{equation}
\Delta p=K_l \frac{\Delta V_l}{V_l}.
\label{eq:negative-pressure}
\end{equation}

We performed this analysis for a total of  thirteen bubbles, see Fig. \ref{fig:Fig3}, showing that the released pressure amounts to values ranging \OV{within the tens of MPa range ($30$ MPa $\pm 16$ MPa).
Due to the measurement error in the image analysis, the estimated pressure presents a lot of uncertainty, but is  comparable to that measured  for cavitation  in  static hydrogels  with the same composition} (22 MPa) \cite{Wheeler2008,Wheeler2009}, and to other experimental methods\cite{Herbert2006}. 
\Ph{Note that the pressure released by triggered cavitation is of the same order than with spontaneous cavitation (Fig 3d).  Indeed, the laser perturbation is not very large, and in order to trigger cavitation we had to wait for the pressure to be close to the spontaneous threshold, i.e. when cavitation occurs spontaneously on nearby cavities. }

From this pressure, the stored elastic energy of water can be evaluated to be $E_p=-\frac{1}{2}\Delta pV_b$. If all this energy was released into kinetic energy, writing $E_k=\frac{3}{2}\rho_l V_b \dot{R_b}^2$ for a bubble in liquid\cite{Leighton1994} of density $\rho_l$, we could expect bubble oscillations with maximum velocity  \Os{$\dot{R}_b=(-\Delta p/3\rho_l)^{1/2}\sim$ 100 m/s, and a characteristic  time $\tau=R_b/\dot{R}_b$ of 0.1 $\mu$s only (for a cavity such that $R_{b}$=10 $\mu$m, $R_{c}$=40 $\mu$m). This time is of the same order than   the observed initial relaxation time of the light intensity signal (0.3 $\mu s$). The system shows oscillations  that are rapidly damped,  probably because of dissipation in the gel surrounding the microcavities, besides viscous dissipation in water or acoustic emissions.}

One may wonder about the content of the cavitation bubble. After cavitation, the bubble quickly fills with water vapor, which has a diffusion coefficient $D_\mathrm{vap}= 2.4\times 10^{-5}$ m$^2$s$^{-1}$. With $R_0$  the initial size of the bubble, the typical vapor filling time is $R_0^2/D_\mathrm{vap}\sim 10^{-5}$ s, \OV{thus longer than initial dynamics}. Then it also probably fills with air, that was dissolved in water and in the hydrogel.  These processes might slightly change the pressure inside the bubble, but cannot account for the continuous post-nucleation expansion of the cavitation bubble.

\begin{figure}
	\begin{center}
	\includegraphics[width=8cm]{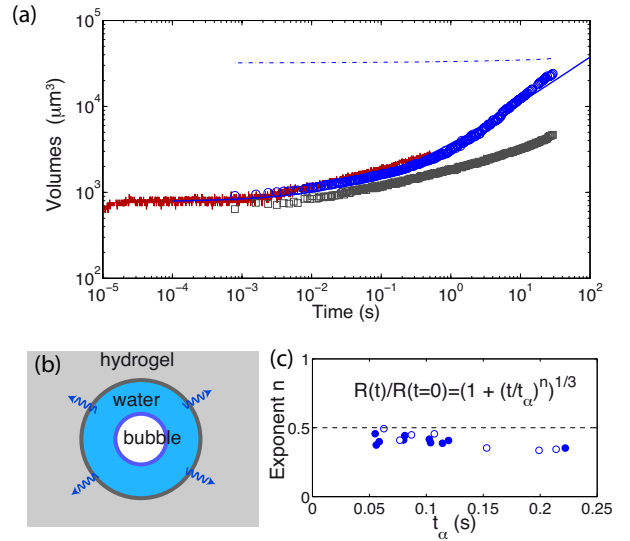}
			\caption{\small \PP{(a)} Long time growth of the bubble volume (red line: \Os{photodiode signal after oscillations}, blue circles: camera) and of the cavity ($V_c$: dashes, $\Delta V_c$ open squares). The bubble volume \Ph{$V_{b}$} \PP{is fitted
			with Eq. \ref{eq:radius-fitting-formula}  (continuous blue line).} 
			(b) Diagram showing the origin of the growth: diffusion of water in the hydrogel. (c) Growth exponent for the bubble size  for triggered (full circles) and spontaneous bubbles (open circles).}
		\label{fig:Fig4}
	\end{center}
\end{figure}

\textit{Long time scales.} Indeed, after its fast expansion \Ph{to the volume $V_{b,0}$}, the bubble is not static but slowly grows, until it completely fills the cavity, after a time of approximately 100 s. To quantify this growth, we analysed the evolution of the volumes of the cavity and of the bubble versus time (Fig. \ref{fig:Fig4}). The bubble expansion \Ph{$V_{b}-V_{b,0}$ versus time}  follows a power law over more than 5 decades of time, with exponents in the range 0.35 to 0.45 (Fig. \ref{fig:Fig4}c).

Neglecting evaporation, there are only two ways to make the bubble grow: increase the volume of the cavity or transfer liquid from the cavity to the gel. The analysis of $V_c(t)$ shows that the size of the cavity indeed increases with time (Fig. \ref{fig:Fig3}c). However, since the bubble eventually fills the whole cavity, an outwards flow of water must occur.

We developed a simple model to account for the power-law expansion of the cavitation bubble. In the following, we will only describe the growth of the bubble by flow in the gel
and hypothesize a constant cavity volume.

When cavitation occurs, the pressure in the liquid suddenly  relaxes from a homogeneous negative value $p_l=-\Delta p$,  to a local value $p_l=p_g-2\sigma/R_0$ with $2\sigma/R_0$ the capillary pressure due to water surface tension, and $p_g$ the gas pressure inside the bubble.
We then have  $\vert p_l \vert \ll \Delta p$. 
This locally changes the osmotic pressure $\Pi=P-p_{l}$ in the gel near the surface of the cavity, where $P$ is the external isotropic stress (here being constant, equal to atmospheric pressure).
Incidently, a gradient of osmotic pressure $\nabla \Pi$  develops across the gel, entailing water fluxes $\mathbf{q}$ which  follow Darcy's law 
\Ph{$\mathbf{q}=-\frac{k(\phi)}{\mu} \nabla p_l=\frac{k(\phi)}{\mu} \nabla \Pi,$} 
with $k$ the permeability, $\mu$ the viscosity, $\phi$ the polymer volume fraction.

The conservation of mass implies\footnote{Conservation of mass for the polymer implies $\phi \,\delta V=\mathrm{cst}$  for an \textit{elementary} hydrogel volume $\delta V$. Its variation is thus: 
$ \frac{d\delta V}{dt}=-\frac{\delta V}{\phi}\frac{d\phi}{dt}$ while conservation of mass for the fluid implies:
$
 \frac{d\delta V}{dt}= \frac{d\delta V_l}{dt}=-\delta V \nabla  \cdot \mathbf{q},$ which gives the formula in the text.
 }:
$
 \frac{\partial\phi}{dt}= \phi \nabla \cdot \mathbf{q}.
 $
A change in volume fraction results in a  change in osmotic pressure, and  the last equation can be rewritten as:
\begin{equation}
\frac{\partial \Pi}{\partial t}= K_\mathrm{sw} \, \nabla . \left(\frac{k}{\mu} \nabla \Pi\right)\;\simeq D \nabla^2 \Pi,
\label{eq:diffusion-equation}
\end{equation}
where $K_\mathrm{sw}=\phi \frac{\partial \Pi}{\partial \phi}$  is the swelling modulus of the gel, describing how the polymer network of the gel swells or shrinks in response to water pressure changes
\footnote{This swelling modulus under constant isotropic stress can also be expressed as $K_\mathrm{sw}= V \frac{\partial p}{\partial V} $, V being the total hydrogel volume. It is equivalent to a compression bulk modulus, but with opposite sign, as a polymer network inflates when increasing the solvent pressure.}.
\Ph{Eq. \ref{eq:diffusion-equation} is similar to the one developed for the sedimentation of a colloidal solution\cite{Kim2007}}.
Assuming low variations of permeability, we  have approximated  the right hand side by
$D \nabla^2 \Pi$. The water content and osmotic pressure therefore follow a diffusion equation, with a diffusion coefficient $D=K_\mathrm{sw}\, k/\mu$.
 The value %
 \footnote{This value is deduced from measurements of the relaxation time of a drying hydrogel. For our pHEMA hydrogel, we estimate $K_\mathrm{sw}\simeq 100$ MPa from Flory's theory \cite{Flory1955}, so the estimated hydrogel permeability is very low: $k/\mu=10^{-19} \mathrm{m}^2/\mathrm{Pa}.\mathrm{s}$. For comparison, with Polyacrilamide hydrogel: Tanaka et al \cite{Tanaka1973} find $D\sim 4 \times 10^{-12}$ m$^2$s$^{-1}$. This diffusion coefficient is  similar to the present pHEMA hydrogel, but originates from very different values of elasticity and permeability, since they have $k/\mu=1/f=5\times 10^{-15} \mathrm{m}^2/\mathrm{Pa}.\mathrm{s}$, but a much softer gel.} %
of this coefficient  is around $D\sim 10^{-11}$ m$^2$s$^{-1}$.

The diffusion of pressure entail fluxes that follow $q= k/\mu \times \Delta p/\sqrt{\pi Dt}$, over the cavity surface $S=4\pi R_c^2$. The transferred volume is therefore $dv_l/dt=q\,S$, so that under the hypothesis of a constant cavity volume $V_{c}$, the bubble volume departs at long times scales from the initial volume $V_0$ (given by Eq. \ref{eq:negative-pressure}) along:
\begin{equation}
\frac{V_b}{V_0}= 1 + \left(\frac{t}{t_\alpha}\right)^n
\label{eq:radius-fitting-formula}
\end{equation}
with $n=1/2$, and 
$t_\alpha= \pi \frac{R_c^2}{D}\left(\frac{K_\mathrm{sw}}{6 \Delta p}  \frac{V_0}{V_c} \right)^{2} ,
\label{eq:talpha}$
which represents the time for which the bubble volume has doubled. 

Fitting experimental curves of bubble growth of Fig. \ref{fig:Fig4}a \OV{(for times t<1s)}, provides \OV{$t_{\alpha}$ and exponent $n$} values displayed on Fig. \ref{fig:Fig4}c. The value of $n$ is close to the predicted value 1/2, except for the largest  characteristic time $t_\alpha$. The observed range of $t_\alpha$ is also in excellent agreement with the expected value, predicted using the negative pressure measured in the first section:
with $\Delta p=20$ MPa, $K_\mathrm{sw}=100$ MPa, $D= 10^{-11}$ m$^2$s$^{-1}$ and values extracted from one experiment of $R_c=23 \mu$m and $R_0=6.8 \mu$m, one finds $t_\alpha=0.08$ s.
The final filling of the bubble, when $R_b=R_c$ can be deduced from Eq. \ref{eq:radius-fitting-formula}, and occurs at a time:
\begin{equation}
t_f= \pi \frac{R_c^2}{D}\left(\frac{K_\mathrm{sw}}{6 \Delta p} \right)^{2},
\label{eq:filling-time}
\end{equation}
where we have neglected $V_0$ in front of $V_c$. 
Using the same numerical values as above, one finds $t_f=115$ s, in very good agreement with the experimental observations. Note that the cavity volume is not constant in \OV{the} experiment, so that  the diffusion model should describe primarily the transferred liquid volume. Preliminary analysis of this volume provides larger growth exponents: this phenomenon is still under investigation.
\OV{Also another challenge is to understand why}, surprisingly, the curves of Fig. \ref{fig:Fig4}a, show that  there is no detectable transfer of fluid to the gel during the first 10 ms after cavitation, since $V_b(t)$ and $\Delta V_c(t)$ initially differ by a constant offset. 
Then, an outward flow proceeds, which results in gas filling of the cavity.

In conclusion, we report in this Letter the first direct observations of spontaneous and triggered cavitation  of water under static tension in confined geometries. The \OV{nucleated} bubble initially expands at a sub-microsecond timescale, driven by a tension of tens of MPa. These extreme tensions result in visible  water volume changes. 
\OV{These extereme tensions lead to uncommon facts in microfluidics: very fast velocities (30 m/s) and visible liquid volume changes associated with inertia, as can be seen on the initial damped oscillations of the bubble. Work is under way to fully characterize these oscillations.}
Also, the cavitation events studied here show a milder dynamic as compared to a free liquid with  successive collapses. Thus, cavitation in water under tension leads to nucleation of bubbles but not to the destructive effects observed during cavitation bubble collapse. 
The presence of hydrogel in trees has been demonstrated recently  \cite{Zwieniecki2001} and could thus have a role in the prevention of damage by cavitation events.
 
We also investigated the following slow growth of the bubble, which is limited by liquid flow in the surrounding gel, rehydrating it. Besides being useful for the design of hydrogel-made microfluidic devices, this fact leads to two remarks. 

First, diffusion in a hydrogel is a complex process: the coefficient $D$ arises from a subtle balance between the elasticity and the permeability of the gel \OV{(Eq. \ref{eq:diffusion-equation})}, which change when it dries or hydrates. Since the hydrogel becomes softer as it rewets, this might slow the diffusion process as the cavitation bubble expands, explaining the fact that the growth exponents seen on Fig. \ref{fig:Fig4} are always lower than expected from normal diffusion. 
Another challenge is to explain the origin of the latency time of about 10 ms after cavitation when no flow towards the gel occurs. 

Second, the release of tension propagates in the gel \OV{when is rewetted by water, and this} effect should be felt in closeby cavities. Indeed, when focusing the laser in the middle of three cavities at the same distance from each one, we noticed that the last triggered cavitation required \OV{much} more laser energy than the first one. 
This interaction might stabilize neighboring cavities, but only for limited times, since diffusion eventually smoothes out the temporary pressure peak.
Such a phenomenon suggest that cavitation propagation can be hindered by negative correlations. Applied to the context of sap transport,  we hypothesize that this phenomenon may play a role in trees to regulate tension in the xylem.

The authors thank Dr. Roberto Gonzalez and Dr. Keita Ando for discussions and help with the experiments and the French-Singapore Merlion grant 2.08.09  for financial support.


\end{document}